\documentclass[aps,prx,reprint,superscriptaddress,showpacs,preprintnumbers,amsmath,amssymb]{revtex4-1}
\usepackage[dvipdfmx]{graphicx}

\usepackage{dcolumn}   
\usepackage{bm}        
\usepackage{amssymb, amsmath, amsfonts}   
\usepackage{color}
\DeclareGraphicsExtensions{{.pdf}}
\begin{document}
\newcommand{\noter}[1]{{\color{red}{#1}}}
\newcommand{\noteb}[1]{{\color{blue}{#1}}}
\newcommand{\field}{\left( \boldsymbol{r}\right)}
\newcommand{\paren}[1]{\left({#1}\right)}
\newcommand{\vect}[1]{\boldsymbol{#1}}
\newcommand{\uvect}[1]{\tilde{\boldsymbol{#1}}}
\newcommand{\vdot}[1]{\dot{\boldsymbol{#1}}}
\newcommand{\vder}{\boldsymbol{\nabla}}

\renewcommand{\phi}{\varphi}
\newcommand{\phiJ}{\varphi_{\rm J}}
\newcommand{\be}{\begin{equation}}
\newcommand{\ee}{\end{equation}}
\newcommand{\bea}{\begin{equnaray}}
\newcommand{\eea}{\end{equnaray}}
\newcommand{\ba}{\begin{align}}
\newcommand{\ea}{\end{align}}
\newcommand{\ave}[1]{\left\langle {#1} \right\rangle}
\newcommand{\tk}[1]{{\color{red}{#1}}}
%
\widetext
%
%
\title{
Shear jamming and shear melting in mechanically trained frictionless particles}
\author{Takeshi Kawasaki}
\affiliation{Department of Physics, Nagoya University, Nagoya 464-8602, Japan}
\author{Kunimasa Miyazaki}
\affiliation{Department of Physics, Nagoya University, Nagoya 464-8602, Japan}

\date{\today}
\begin{abstract}
We investigate criticality near the jamming transition in both quiescent
 systems and those under shear by considering the effect of mechanical
 training on the jamming transition and nonlinear rheology. We simulate
 frictionless soft particles undergoing athermal quasi-static shear
 using initial configurations trained with athermal quasi-static cyclic
 volume deformations. The jamming transition density of the initial
 configuration $\varphi_{\rm J0}$ is systematically altered by tuning
 the ``depth'' of mechanical training. We exert a steady shear on these
 configurations and observe either shear jamming (gain of stiffness due
 to shear) or shear melting (loss of stiffness due to shear), depending
 on the depth of training and proximity to the jamming transition
 density. We also observe that the characteristic strains, at which
 shear jamming or melting occur, diverge at a unique density
 $\varphi_{\rm JS}$. This is due to the shift of the jamming transition
 density from $\varphi_{\rm J0}$ to $\varphi_{\rm JS}$ under shear,
 associated with loss of memory of the initial configuration. Finally,
 we thoroughly investigate nonlinear rheology near the jamming
 transition density, and contrary to previous works, we find a nonlinear
 ``softening'' takes place below as well as above the jamming transition density.
\end{abstract}
\maketitle
%
\section{Introduction}
A disordered packing of grains becomes rigid when its density exceeds the jamming transition density $\varphi_{\rm J}$~\cite{Liu1998Nature}. In the vicinity of $\varphi_{\rm J}$, critical behavior is observed for various mechanical quantities; examples include elastic moduli, pressure, and yield stress~\cite{OHern2002Phys.Rev.Lett.,OHern2003Phys.Rev.E,Makse1999Phys.Rev.Lett.,Olsson2007Phys.Rev.Lett.,Otsuki2009Phys.Rev.E,Hecke2010J.Phys.:Condens.Matter,Kawasaki2015Phys.Rev.E,Vagberg2016Phys.Rev.E}. Moreover, the mechanical response near $\varphi_{\rm J}$ is highly nonlinear and complex. 
Recent studies have shown that at a density slightly above $\varphi_{\rm J}$, the stress-strain curve shows ``softening'' in which the shear stress $\sigma$ becomes hypo-elastic and is proportional to $\sqrt{\gamma}$ in a small strain regime following the linear elastic regime~\cite{Coulais2014Phys.Rev.Lett.,Otsuki2014Phys.Rev.E,Nakayama2016J.Stat.Mech.,Boschan2016SoftMatter,Dagois-Bohy2017SoftMatter}. It is also claimed that the onset strain at which softening occurs depends on proximity to the jamming transition density $\delta \phi=\phi-\phi_{\rm J}$, and controversially, its critical exponent has been reported as being 0.75~\cite{Nakayama2016J.Stat.Mech.,Goodrich2016Proc.Natl.Acad.Sci.USA} or 1.0~\cite{Otsuki2014Phys.Rev.E,Boschan2016SoftMatter,Dagois-Bohy2017SoftMatter}. Moreover, its physical mechanism remains elusive.

For larger strains, the stress is known to become constant due to the
incidence of macroscopic plastic events. This stress is called the yield
stress $\sigma^{\rm Y}$ and is believed to obey a critical behavior on
approaching the jamming transition
density~\cite{Olsson2007Phys.Rev.Lett.,Otsuki2009Phys.Rev.E,Vagberg2016Phys.Rev.E}. In
previous studies, the quasi-static limit of $\sigma^{\rm Y}$ was
obtained using the Herschel-Bulkley (HB) law ($\sigma =
\sigma^{\rm Y}+A\dot{\gamma}^B$), derived from how the shear stress
$\sigma$ varied with strain rate in a finite shear rate
system~\cite{Larson1999,Hohler2005J.Phys.:Condens.Matter,Olsson2012Phys.Rev.Lett.,Ikeda2012Phys.Rev.Lett.,Dinkgreve2015Phys.Rev.E,Bonn2017Rev.Mod.Phys.,Tighe2010Phys.Rev.Lett.,Olsson2011Phys.Rev.E,Hatano2011J.Phys.:Conf.Ser.}. However,
for finite shear rate simulations/experiments, it is known that
obtaining the yield stress near $\phi_{\rm J}$ is difficult, since the
infinitely small shear rates are required. This might be a source of
contention for determining the critical exponent of $\sigma^{\rm Y}$
with respect to $\delta \phi$, the proximity to the jamming transition
density~\cite{Otsuki2009Phys.Rev.E,Tighe2010Phys.Rev.Lett.,Olsson2011Phys.Rev.E,Hatano2011J.Phys.:Conf.Ser.,Vagberg2016Phys.Rev.E}:
the exponent varies in the range of [1.0, 1.5] for harmonic potential
systems. In order to obtain the yield stress in an asymptotic, athermal
quasi-static (AQS) state, another simulation technique has been used,
where successive discrete shear strains $\Delta \gamma$ are applied with
energy minimization, {\it i.e.}, the system is always at a local
minimum of the energy landscape. Even with this approach, different 
values of the critical exponent for the yield stress are reported for
harmonic potential
systems~\cite{Heussinger2010SoftMatter,Hatano2011J.Phys.:Conf.Ser.}.
The critical behavior of the yielding stress remains elusive and 
a new approach is required.

The jamming transition density $\varphi_{\rm J}$ is known to be strongly dependent on preparation protocols for jammed configurations~\cite{Chaudhuri2010Phys.Rev.Lett.,Ozawa2012Phys.Rev.Lett.,Kumar2016GranularMatter}. It is possible to change $\varphi_{\rm J}$ systematically by exposing the system to thermal fluctuations or mechanical deformations, so-called ``thermal annealing"~\cite{Chaudhuri2010Phys.Rev.Lett.,Ozawa2012Phys.Rev.Lett.} or ``mechanical training"~\cite{Kumar2016GranularMatter}, respectively. 
Recently it has been found that applying shear strain below $\phi_{\rm
J}$ triggers shear jamming, {\it i.e.}, acquiring rigidity by applying
shear strain. 
Shear jamming has been observed for mechanically trained frictionless particles~\cite{Kumar2016GranularMatter}. 
Given that it has been commonly believed, until recently, that shear
jamming could only be observed in systems composed of the particles with
frictional contacts~\cite{Bi2011Nature}, it is striking that the
the unjammed packing can undergo the shear jamming in the absence of the
friction as long as the packing configurations are generated
using proper training or annealing. 
However, the whole pictures of shear jamming and accompanied nonlinear
rheological behaviors are yet to be elucidated.

In this study, we focus on changes in the
jamming transition density when shear is applied to a mechanically
trained configuration. A shifted jamming transition may account for an
unprecedented behavior of nonlinear rheological phenomena within the
same framework. For example, by using a well-trained configuration,
shear is expected to lead to loss of memory of the initial
configuration. The structure will become disordered, resulting in a
decrease in the jamming transition density. If this is the case, by
using an initial configuration slightly below the jamming
transition density, we will observe an unjammed to jammed transition, 
{\it i.e.}, shear jamming. In a less trained configuration, the jamming
transition density is not significantly altered; thus, shear jamming is
not observed when shear is applied. For very poorly trained
configurations, the jamming transition density actually increases. In
this case, a transition from jammed to unjammed states, or shear
melting,  takes place. 
We seek to resolve the mechanism behind different instances of nonlinear rheology observed near the jamming transition by systematically tuning the degree of mechanical training of the initial configuration following the protocol proposed by Ref.~\cite{Kumar2016GranularMatter}.
 
Firstly, we describe the simulation methods and how mechanically trained initial configurations are generated. 
Next, we discuss how the jamming transition density varies depending on the depth of mechanical training. We then go on to examine the mechanical response of configurations with different depths of mechanical training. Furthermore, we demonstrate the mechanism behind the complex mechanical responses of these packings by focusing on the development of the jamming transition density when shear is applied. Finally, we discuss the critical behavior of both static and dynamic quantities in the athermal quasi-static limit. 

\begin{figure}
\includegraphics[width=8cm]{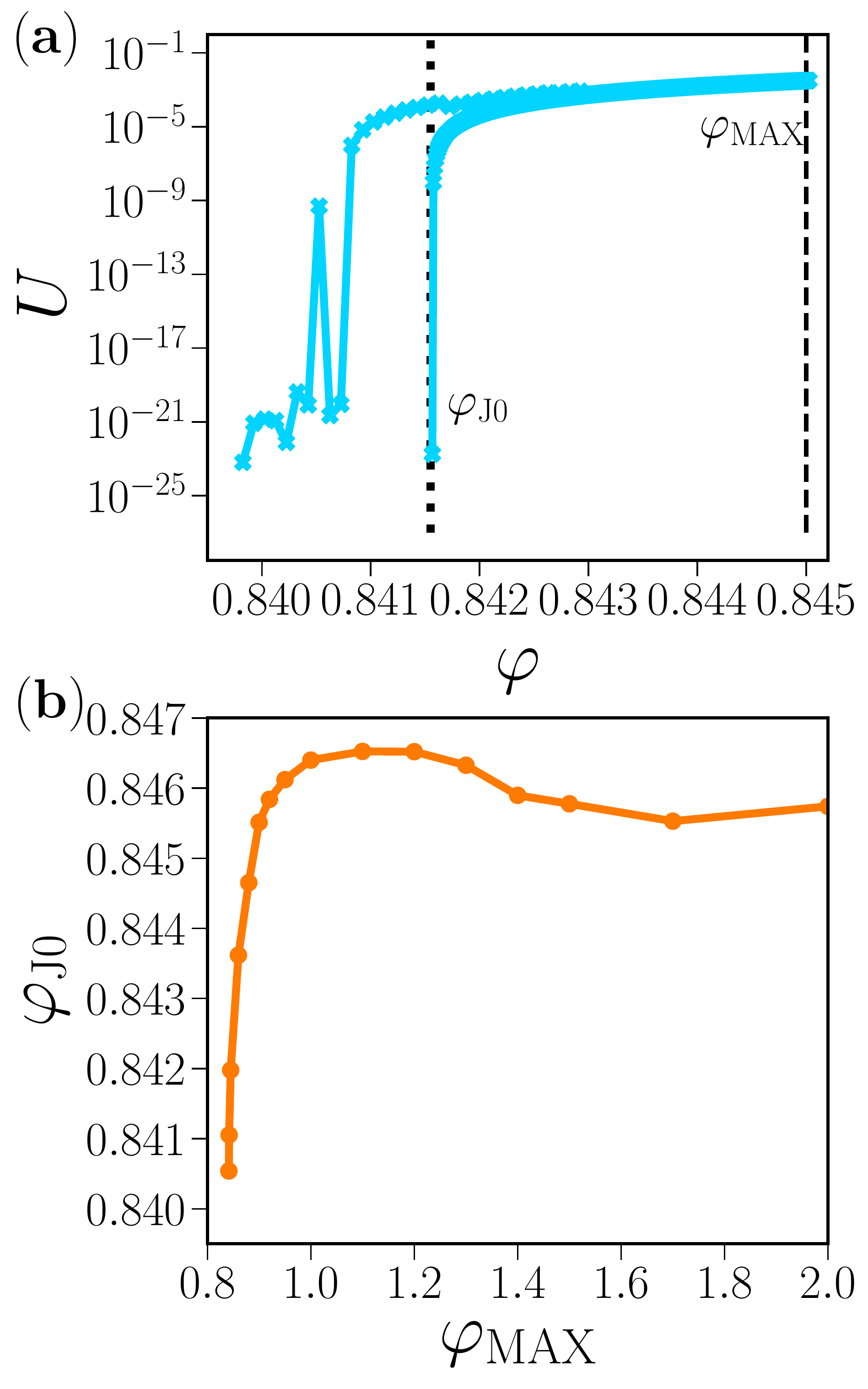}
\caption{(Color online) 
(a) Preparation protocol for initial configurations: $\phi$ dependence of potential energy per particle $U$ during the compression-decompression process. $\phi_{\rm MAX}$ is the maximum packing fraction during this process, and $\phi_{J0}$ is the packing fraction when $U<10^{-16}$ for the first time during decompression. The cross symbols represent the state points where we carry out energy minimizations. (b) Jamming transition density, $\phi_{\rm J0}$ as a function of the depth of the mechanical training, $\phi_{\rm MAX}$. As $\phi_{\rm MAX}$ is increased, $\phi_{\rm J}$ increases until $\phi_{\rm MAX}\sim 1.2$. When $\phi_{\rm MAX}$ is above 1.2, $\phi_{\rm J0}$ slightly decreases and converges to $\phi_{\rm J0}\sim 0.846$. 
\label{fig:fig1}}
\end{figure}

\section{Numerical modeling}
The system we study is a two-dimensional equimolar binary mixture of
frictionless particles with diameters $\sigma_{\rm L}$ and $\sigma_{\rm
S}$. The size ratio of small and large particles is $\sigma_{\rm
L}/\sigma_{\rm S}= 1.4$. The particle number is $N=1156$ unless
otherwise stated. To investigate the finite size effect, simulations ofdifferent sizes are also performed in the range $N =$ 288 to 3538. This
is provided as Supplementary Information~\cite{zotero-12580} (see
Fig.~S5) and it is confirmed that the finite size effect
does not affect the main results. The interaction~\cite{Durian1995Phys.Rev.Lett.} between the $j$-th and $k$-th particles is 
the harmonic potential defined by 
\be
U(r_{jk})=\frac{\epsilon}{2}\left\{1-(r_{jk}/\sigma_{jk})\right\}^2, 
\ee
where $r_{jk} = |{\bf r}_j-{\bf r}_k|$ and $\sigma_{jk} =
(\sigma_j+\sigma_k)/2$. Here $\sigma_{j(k)}$ is the diameter of the
$j(k)$-th particle. In our simulations, we use $\sigma_{\rm S}$,
$\epsilon$, and $\epsilon/\sigma_{\rm S}^2$ as units of length, energy
and stress, respectively. The particles are driven to a quasi-static
state by employing the FIRE algorithm~\cite{Bitzek2006Phys.Rev.Lett.}
for energy minimization. We also apply shear stabilization to remove any
residual stress for the initial
configurations~\cite{Dagois-Bohy2012Phys.Rev.Lett.,Shuang2019ComputationalMaterialsScience}. A
description of the FIRE algorithm with shear stabilization is provided
in Supplementary Information~\cite{zotero-12580}. We consider a configuration to be quasi-static when the average force amplitude acting on a particle is less than $10^{-14}\epsilon/\sigma_{\rm S}$. This threshold value is determined by the numerical accuracy of double precision numbers, plus round-off errors due to the summation of forces on neighboring particles.  We perform constant volume simulations for the most part; an exceptions are the data shown in Fig.~\ref{fig:fig5}~(b) and Supplementary Information obtained from quasi-static constant pressure simulations. Details are provided in Supplementary Information~\cite{zotero-12580}.


\begin{figure}
\includegraphics[width=\linewidth]{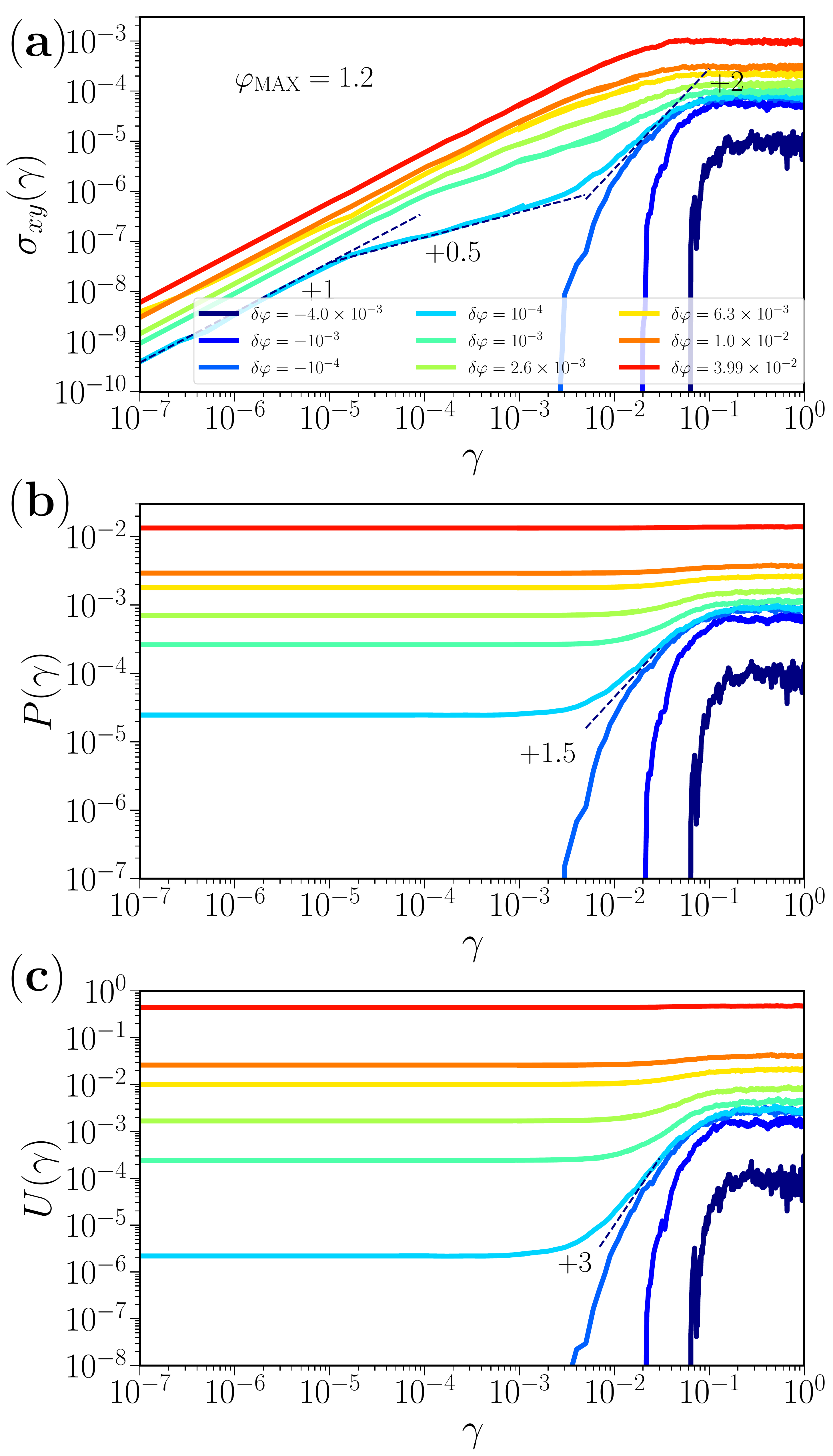}
\caption{Mechanical response to strain $\gamma$ for various $\delta \phi
 (=\phi-\phi_{\rm J0})$. Initial configurations are prepared with
 $\phi_{\rm MAX}=1.2$. (a) $\sigma_{xy}$ vs $\gamma$. When $\delta
 \phi\gtrsim0$, as $\gamma$ is increased, the stress-strain curves show
 an elastic response $\sigma_{xy}=G\gamma$ at very small $\gamma$, where
 $G$ is the shear modulus. At intermediate $\gamma$, we observe
 softening behavior $\sigma_{xy}\sim \gamma^{1/2}$, another instance of
 nonlinear response. At even larger $\gamma$, we see shear hardening,
 $\sigma_{xy}\sim \gamma^2 $, followed by yielding ($\sigma_{xy} \sim$
 constant). When $\delta \phi\lesssim0$, the stress-strain curves show
 shear jamming behavior, {\it i.e.},  $\sigma_{xy}\sim 0$ at small $\gamma$, but becomes non-zero for intermediate $\gamma$. This is followed by a regime where $\sigma_{xy} \sim \gamma^2$. At even larger $\gamma$, it yields. (b) $P$ vs $\gamma$. When $\delta \phi>0$, as $\gamma$ is increased, $P$ is almost constant over the elastic and softening regimes. $P\sim \gamma^{1.5}$ in the shear hardening regime. When $\delta \phi<0$, shear jamming behavior is obtained, similar to $\sigma_{xy}$ vs $\gamma$. (c) $U$ vs $\gamma$. This is similar to $P$ vs $\gamma$ except for the slope of the shear hardening regime ($U\sim \gamma^3$).
\label{fig:fig2}}
\end{figure}
\section{Results}
\subsection{Computing jamming configurations}
We produced initial configurations using a quasi-static cyclic volume deformation; this corresponds to mechanical training. Though this is equivalent to what was introduced in Ref.~\cite{Chaudhuri2010Phys.Rev.Lett.,Kumar2016GranularMatter}, the present study uses a wider range of ``depths'' of mechanical training compared to previous work~\cite{Kumar2016GranularMatter}. ``Depth'' here is defined as the maximum density $\phi_{\rm MAX}$ to which the system is compressed during the cyclic deformation.  As shown in Fig.~\ref{fig:fig1}~(a), we firstly prepare a random configuration at $\phi = 0.8395$ and increase $\phi$ in $10^{-4}$ steps until $\phi_{\rm MAX}$. Subsequently, we decrease $\phi$ in $10^{-4}$ steps if $U>10^{-8}$, otherwise in $10^{-6}$ steps. We note that with the default system size ($N=1156$), $\Delta\phi = 10^{-6}$ is the smallest meaningful increment; smaller steps may not be applied due to finite size effects~\cite{Goodrich2012Phys.Rev.Lett.,Goodrich2014Phys.Rev.E}. When the potential energy becomes $U<10^{-16}$ for the first time, we define the corresponding packing fraction to be $\phi_{\rm J0}$ (see Fig.~\ref{fig:fig1}~(a) when $\phi_{\rm MAX}=0.845$). It was found that the jamming transition density $\phi_{\rm J0}$ varies non-monotonically with training depth. Figure~\ref{fig:fig1}~(b) shows $\phi_{\rm J0}$ as a function of $\phi_{\rm MAX}$. We can see that as $\phi_{\rm MAX}$ increases, $\phi_{\rm J0}$ also increases when $\phi_{\rm MAX}$ is less than 1.2; when $\phi_{\rm MAX}$ is greater than 1.2, $\phi_{\rm J0}$ slightly decreases and then converges to $\phi_{\rm J0}\sim 0.846$.
\begin{figure}
\includegraphics[width=\linewidth]{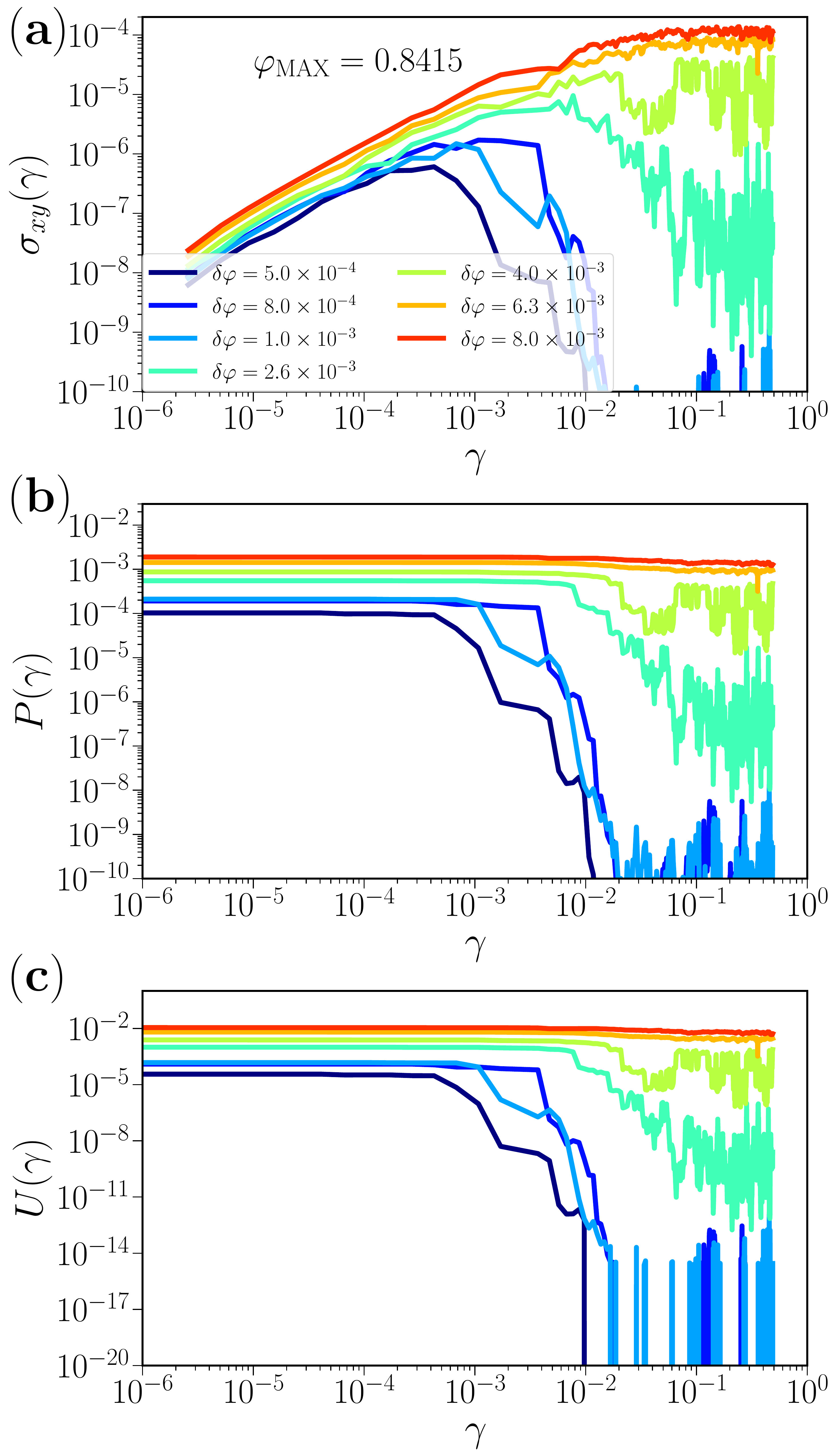}
\caption{Mechanical response to strain $\gamma$ for various $\delta \phi (=\phi-\phi_{\rm J0})$. Initial configurations are prepared with $\phi_{\rm MAX}=0.8415$. (a) $\sigma_{xy}$ vs $\gamma$. Note that when $\delta \phi$ is small, $\sigma_{xy}$ drops to zero at intermediate $\gamma$, indicating shear melting. (b) and (c) $P$ vs $\gamma$ and $U$ vs $\gamma$ respectively. When $\delta \phi$ is small, $P$ and $U$ also exhibit shear melting at intermediate $\gamma$.
\label{fig:fig3}}
\end{figure}
\begin{figure}
\includegraphics[width=\linewidth]{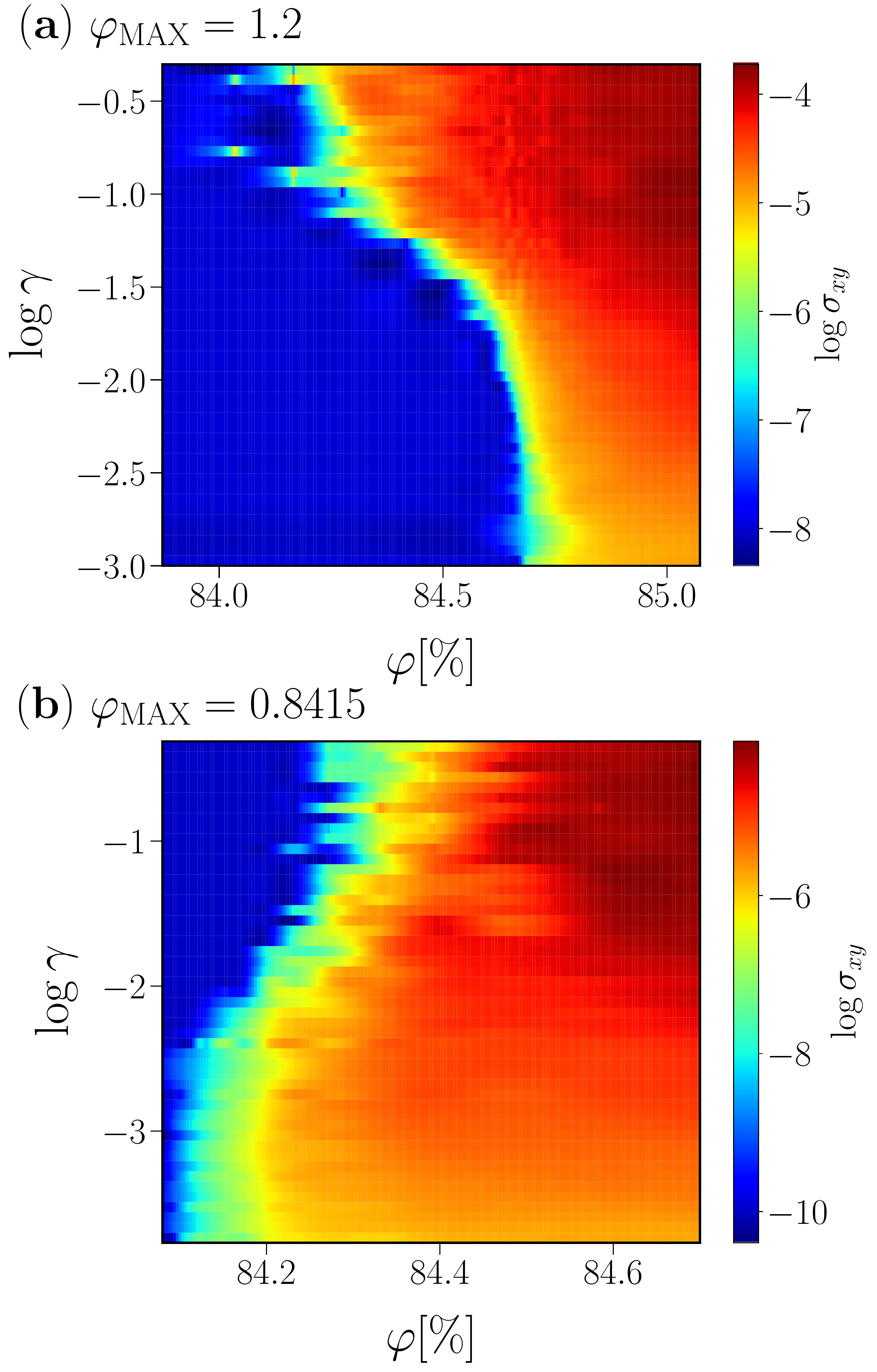}
\caption{(a) Shear stress $\sigma_{xy}$ color map for $\phi_{\rm
 MAX}=1.2$, $\phi_{\rm J0}=0.8465$ for different shear strains $\gamma$
 and packing fractions $\phi$. When $\phi\lesssim\phi_{J0}$, as $\gamma$
 increases, $\sigma_{xy}$ becomes non-zero at a finite $\gamma$. This is
 indicative of shear jamming. Shear jamming does not take place when
 $\phi \lesssim 0.843$. When $\phi\gtrsim\phi_{J0}$, the system always
 shows jamming behavior, {\it i.e.}, $\sigma_{xy}$ is positive and
 non-zero. (b) Shear stress $\sigma_{xy}$ color map for $\phi_{\rm
 MAX}=0.8415$, $\phi_{\rm J0}=0.8405$ for different shear strains
 $\gamma$ and packing fractions $\phi$. When $\phi\lesssim 0.843$, as
 $\gamma$ increases, $\sigma_{xy}$ becomes zero, {\it i.e.}, unjams at $\gamma \sim10^{-2}$. This is indicative of shear melting. When $\phi\gtrsim 0.843$, shear melting does not take place, and the system always shows jamming behavior. 
\label{fig:fig4}}
\end{figure}
\subsection{Mechanical response}
We firstly consider the mechanical response of these configurations to quasi-static steady shear~\cite{Heussinger2009Phys.Rev.Lett.} using Lees-Edwards boundary conditions~\cite{Allen1988}. With each step, a small shear affine strain is applied to drive the particles in the shear direction by  
\be
{\bf r}'_j(n+1) = {\bf r}_j(n) + \Delta \gamma(n) y_j(n) {\bf e}_x,
\ee
where ${\bf r}_j(n)$ is the position of the $j$-th particle at the
$n$-th simulation step. After each step, the positions of the particles
${\bf r}'_j(n+1)$ are relaxed using the FIRE algorithm to minimize the
energy. The shear strain evolves as $\gamma(n+1) = \gamma(n)+
\Delta\gamma(n)$. When the accumulated shear strain is in the regime
$\gamma<10^{-3}$, $\Delta \gamma(n)$ is logarithmically increased from
$10^{-7}$ (or $10^{-9}$) to $10^{-3}$; when $\gamma>10^{-3}$, $\Delta \gamma(n) =
10^{-3}$. The shear stress, normal stress (or the pressure), and total potential energy are measured using a quasi-static steady shear configuration. 
The stress tensor is defined as
\be
\sigma_{\alpha\beta}
=\frac{1}{2L^2}\sum_{j,k}\frac{r^{\alpha}_{jk}r^{\beta}_{jk}}{r_{jk}^2}\frac{\partial U}{\partial r_{jk}},
\label{eq:virial}
\ee
where $\alpha, \beta \in \{x,y\}$, $r_{jk}^{x}=x_{jk}$ and $r^{y}_{jk}=y_{jk}$. The shear stress is given by the off-diagonal components of the stress tensor, $\sigma_{xy}$ or $\sigma_{yx}$. The pressure is calculated from the diagonal components, $P=-(\sigma_{xx}+\sigma_{yy})/2$. The potential energy per particle is found from 
\be
U=\frac{1}{2N}\sum_{j,k}U({\bf r}_{jk}).
\ee

We consider the mechanical response to quasi-static shear of the initial configurations that are
mechanically trained with different $\phi_{\rm MAX}$. All data shown
below are averaged over at least 15 independent runs (typically, more
than 50 runs). Figure~\ref{fig:fig2} shows the response as a function of
$\gamma$ for various $\delta \phi (=\phi-\phi_{\rm J0})$ when $\phi_{\rm
MAX}=1.2$. Figure~\ref{fig:fig2} (a) shows the $\gamma$ dependence of
the shear stress $\sigma_{xy}$, or the stress-strain curves. Slightly
above the jamming transition, $\delta \phi \gtrsim 0$, the stress-strain
curve exhibits a unique behavior as the shear strain is increased. For
small $\gamma$, we see an elastic response $\sigma_{xy}=G\gamma$, where
$G$ is the shear modulus. At intermediate $\gamma$'s,  
following the elastic regime, a nonlinear behavior, which is called
``shear softening''  is observed, where $\sigma_{xy}\sim \gamma^{1/2}$. 
At larger $\gamma$'s, the stress increases sharply as $\sigma_{xy}\sim
\gamma^2$, which we shall refer to as the ``shear hardening''.
At even large $\gamma$ beyond this hardening regime, the system
eventually yields and $\sigma_{xy}$ becomes constant. 
Here, we find that the characteristic shear strain for the onset of
softening $\gamma_{\rm s}$ depends on $\delta \phi$, as observed in
Ref.~\cite{Boschan2016SoftMatter}. This will be discussed later. When
$\delta \phi\lesssim0$, on the other hand, the stress-strain curves show
shear jamming behavior, {\it i.e.}, $\sigma_{xy}\sim 0$ at small $\gamma$, $\sigma_{xy} \sim \gamma^2$ at intermediate $\gamma$, and constant at large $\gamma$ (yielding).

\begin{figure}
\includegraphics[width=\linewidth]{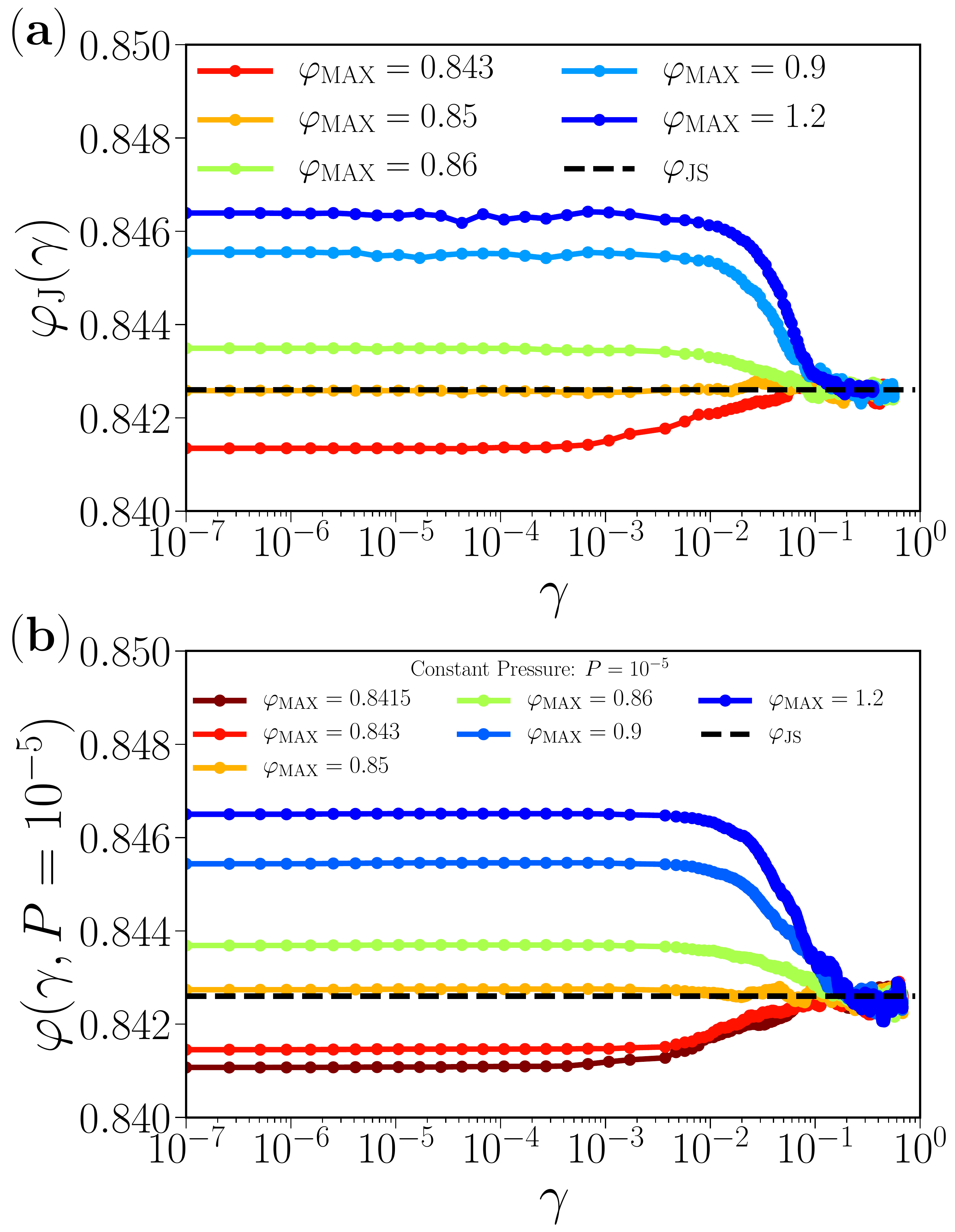}
\caption{(a)Jamming transition density as a function of $ \gamma $, $\phi_{\rm J}(\gamma)$ for various $\phi_{\rm MAX}$. $\phi_{\rm J}(\gamma)$ deviates from $\phi_{J0}$ and converges at large $\gamma$ to another characteristic density $\sim 0.8426$, called $\phi_{\rm JS}$, the jamming transition point for sheared configurations. The dash line is $\phi_{\rm J}(\gamma)=\phi_{\rm JS}$. (b) The packing fraction $\phi$ obtained from constant pressure simulations at very small pressure $P=10^{-5}$ with shear strain $\gamma$ for configurations trained with various $\phi_{\rm MAX}$. $\phi(\gamma,P=10^{-5})$ is approximately identical to $\phi_{\rm J}(\gamma)$.
\label{fig:fig5}}
\end{figure}

\begin{figure}
\includegraphics[width=8.5cm]{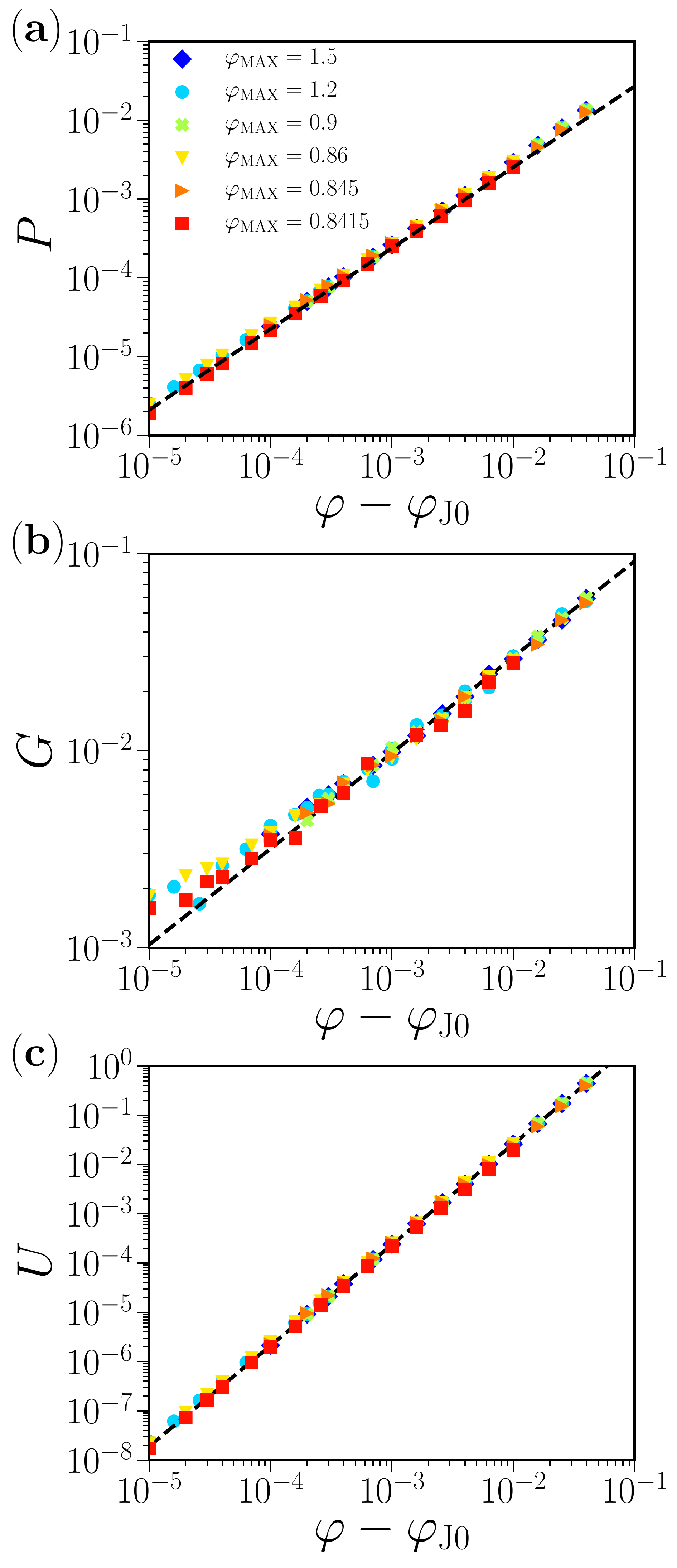}
\caption{
Critical scaling of the mechanical properties of initial configurations trained with various $\phi_{\rm MAX}$. (a)~Dependence of pressure $P$ on $\phi-\phi_{\rm J0}$. The dashed line is a power law fit, $P=A(\phi-\phi_{\rm J0})^{\alpha}$, where $A=0.2866$ and $\alpha=1.0266$. (b)~Dependence of the shear modulus $G$ on $\phi-\phi_{\rm J0}$. The dashed line is a power law fit, $G=B(\phi-\phi_{\rm J0})^{\beta}$, where $B=0.2811$ and $\beta=0.487$.  (c)~Dependence of the potential energy $U$ on $\phi-\phi_{\rm J0}$. The dashed line is a power law fit, $U=C(\phi-\phi_{\rm J0})^{\gamma}$, where $C=325.14$ and $\gamma=2.04703$. 
\label{fig:fig6}}
\end{figure}

\begin{figure}
\includegraphics[width=8.5cm]{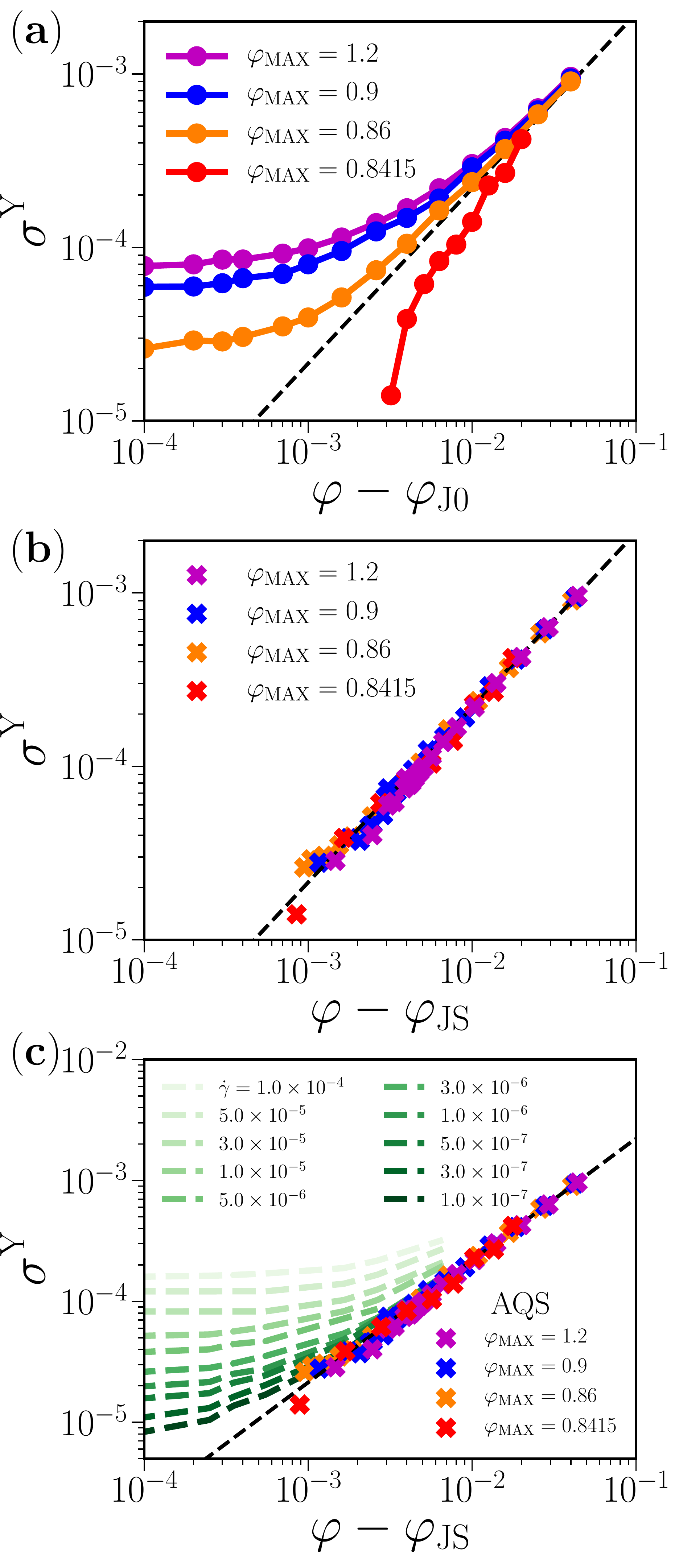}
\caption{(a) Yield stress $\sigma^{\rm Y}$ as a function of $\phi-\phi_{\rm J0}$ (shown as filled circles) for different training depths $\phi_{\rm MAX}$. They show deviations from critical behavior; the degree of deviation depends on the degree of training $\phi_{\rm MAX}$. (b) Yield stress $\sigma^{\rm Y}$ as a function of $\phi-\phi_{\rm JS}$ (shown as cross marks) for various $\phi_{\rm MAX}$. We use $\phi_{\rm JS}=0.8426$ for all the data. $\sigma^{\rm Y}$ vs $\phi-\phi_{\rm JS}$ shows critical behavior for any $\phi_{\rm MAX}$, satisfying $\sigma^{\rm Y}=A(\phi-\phi_{\rm JS})^{1.01}$. (c) Steady state shear stress obtained using finite shear rate simulations for various shear rates $\dot{\gamma}$ and packing fractions $\phi$, reproduced from Ref.~\cite{Vagberg2016Phys.Rev.E}. Compared with plots of $\sigma^{\rm Y}$ vs $\phi-\phi_{\rm JS}$ from our AQS simulations.
\label{fig:fig7}}
\end{figure}

\begin{figure}
\includegraphics[width=8.5cm]{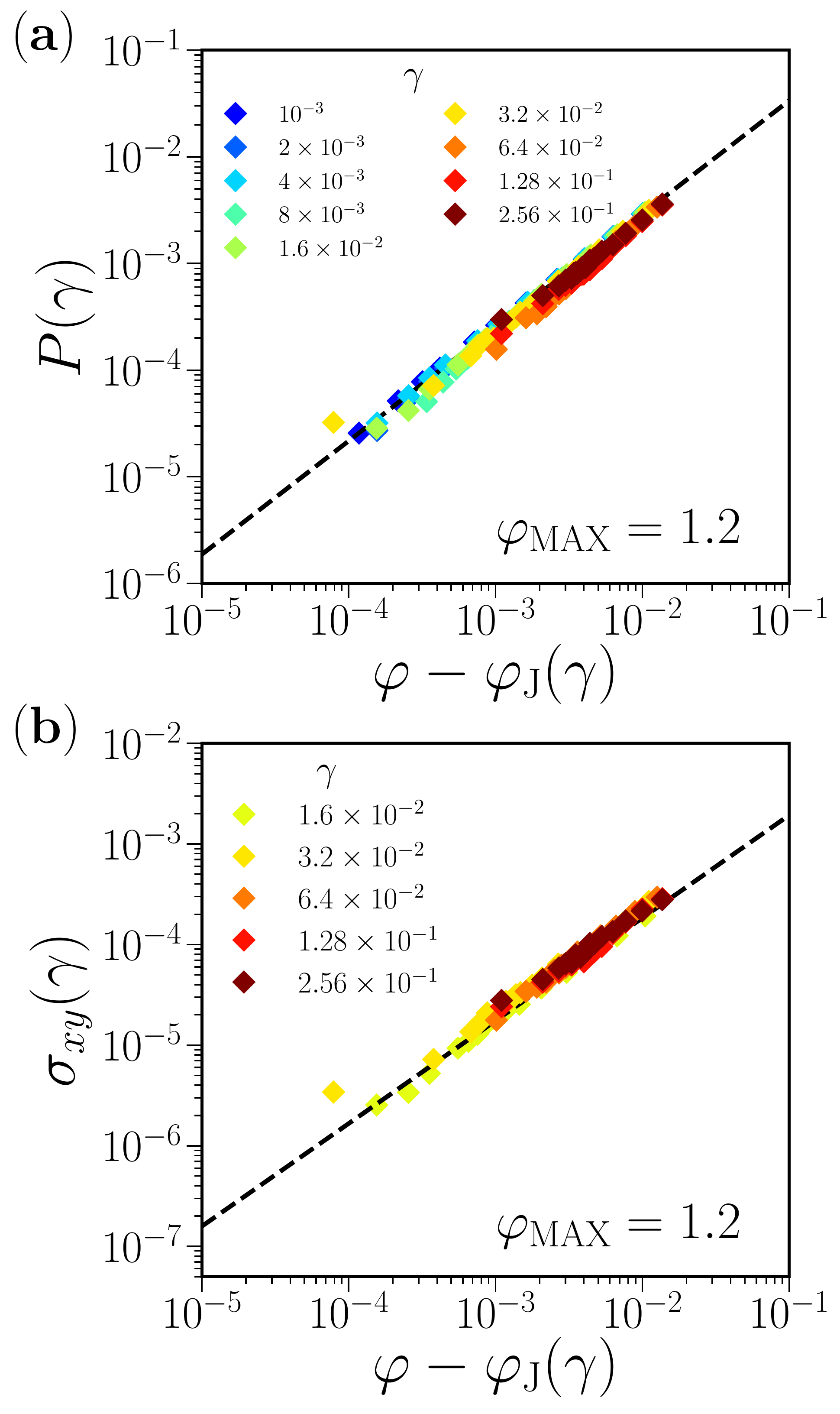}
\caption{(a) Pressure $P$ as a function of $\phi-\phi_{\rm J}(\gamma)$, where strain $\gamma$ is applied to configurations trained with $\phi_{\rm MAX}=1.2$. $\phi_{\rm J}(\gamma)$ values are the same as those obtained in Fig.~\ref{fig:fig5}. Over a wide range of $\gamma$, all the data collapses along $P \sim P_0(\phi-\phi_{J}(\gamma))^{1.05}$, where $P_0=0.3885$. (b) Shear stress $\sigma_{xy}$ as a function of $\delta \phi=\phi-\phi_{\rm J}(\gamma)$, where a strain $\gamma$ is applied to configurations trained with $\phi_{\rm MAX}$=1.2. When $\gamma \gtrsim 0.01$, the data collapses along $\sigma_{xy}(\gamma) \sim \sigma_0(\phi-\phi_{\rm J}(\gamma))^{1.01}$, where $\sigma_0=0.0203$.
\label{fig:fig8}}
\end{figure}

Figure~\ref{fig:fig2} (b) shows the $\gamma$ dependence of the pressure $P$. This is similar to the stress-strain curves except for the elastic and softening regimes. When $\delta \phi>0$, as $\gamma$ is increased, $P$ is almost constant through both elastic and softening regimes, while it obeys $P\sim \gamma^{1.5}$ in the shear hardening regime. 
Note that the power law exponents for $\sigma_{xy}$ and $P$ with respect
to $\gamma$ are shifted by 0.5 in these regimes. This is attributed to
how their ratio, the friction coefficient $\mu\equiv \sigma_{xy}/P$, varies as $\gamma^{0.5}$. 
This implies 
that the softening regime spreads over a wide range of $\gamma$ near
jamming. This will be discussed further below and in
Fig.~\ref{fig:fig9}. When $\delta \phi\lesssim 0$, shear jamming
behavior is obtained, similar to what we see in the stress-strain
curves. Again, $P$ is not sensitive to the elastic regime nor the
softening behavior. Below, we find that the $\phi$ dependence of the
pressure $P(\gamma, \phi)$ at any shear strain shows critical behavior
when plotted against proximity to the jamming transition density, $\phi_{\rm J}(\gamma)$, at each
corresponding $\gamma$, which will be discussed further in Figs.~\ref{fig:fig5}-~\ref{fig:fig8}. 
Finally, Figure~\ref{fig:fig2} (c) shows how the potential energy $U$ varies with $\gamma$. For all $\phi$, $U$ vs $\gamma$ is similar to $P$ vs $\gamma$ except for the slope of the shear-hardening regime, where $U \sim \gamma^3$. Note that the exponent is double that of $P$. This is due to the relationship $U(\gamma,\phi) \sim \delta \phi(\gamma) \sim  P(\gamma,\phi)^2$ in the case of harmonic interactions.

We also consider mechanical response at different depths of mechanical training. Figures~S1 and~S2 in the Supplementary Information~\cite{zotero-12580} present the mechanical response when $\phi_{\rm MAX}=0.9$ and $\phi_{\rm MAX}=0.86$. We find that the elastic and softening behaviors are identical to what we obtained in Fig.~\ref{fig:fig2} where $\phi_{\rm MAX}=1.2$, 
though the yield stress is different.  
Figs.~\ref{fig:fig6} (a)-(c) shows how the pressure $P$, shear modulus
$G$, and potential energy $U$ depend on $\phi-\phi_{\rm J0}$. Note that
the shear modulus is obtained from the slope of the stress-strain curve, $G={\rm d}\sigma_{xy}(\gamma)/{\rm d}\gamma|_{\gamma=10^{-7}}$. We see that these linear response properties all scale with $\phi-\phi_{\rm J0}$. On the other hand, even for the same $\phi-\phi_{\rm J0} =10^{-4}$, $\sigma^{\rm Y}$ at $\phi_{\rm MAX}=0.86$ is smaller than that at $\phi_{\rm MAX}=1.2$. 
It indicates that the criticality of $\sigma^{\rm Y}$ is different from static mechanical properties e.g.  $P$, $G$, and $U$.

With much less trained configurations, we find significantly different behavior. In fact, we observe shear melting, a behavior which is absent from our well-trained configurations. In Fig.~\ref{fig:fig3}, we show the mechanical response as a function of $\gamma$ for various $\delta \phi (=\phi-\phi_{\rm J0})$ at $\phi_{\rm MAX}=0.8415$. Figure~\ref{fig:fig3}~(a) shows how $\sigma_{xy}$ varies with $\gamma$. 
When $\delta \phi > 0$ but small, 
the stress-strain curves show elastic behaviors for small $\gamma$
followed by a onset of the softening, similar to what we observed for larger $\phi_{\rm MAX}$. 
At intermediate $\gamma$, however, $\sigma_{xy}$ suddenly drops to zero.
This is the shear melting, {\it i.e.},  the transition from jammed to
unjammed states. 
Note that shear melting does not take place when $\delta \phi$ is large. 
Figures~\ref{fig:fig3}~(b) and (c) show $P$ and $U$ as a function of $\gamma$. Both also exhibit shear melting at intermediate $\gamma$.

Finally, we combine the stress-strain curves for a wide-range of
densities when $\phi_{\rm MAX}=1.2$ (well trained) and  $0.8415$ (poorly
trained) into two color maps of the shear stress $\sigma_{xy}$ as a
function of packing fraction $\phi$ and shear strain
$\gamma$. Fig.~\ref{fig:fig4} (a) corresponds to $\phi_{\rm MAX}=1.2$;
note that $\phi_{\rm J0}=0.8465$. When $\phi\lesssim\phi_{J0}$,
$\sigma_{xy}$ becomes non-zero, {\it i.e.}, jams at $\gamma \sim10^{-2}$ with
increasing $\gamma$. This corresponds to shear jamming. This is not the
case when $\phi \lesssim 0.843$, as the configurations unjam under any
shear strain. When $\phi\gtrsim\phi_{J0}$, the system always shows
jamming behavior, where the $\sigma_{xy}$ is positive and
non-zero. Fig.~\ref{fig:fig4} (b) shows the same information for
$\phi_{\rm MAX}=0.8415$ for which $\phi_{\rm J0}=0.8405$. 
When
$\phi\lesssim 0.843$, $\sigma_{xy}$ becomes zero, {\it i.e.}, jams at $\gamma
\sim10^{-2}$ with increasing $\gamma$. This corresponds to shear
melting. When $\phi\gtrsim 0.843$, shear melting does not take place and
the system always shows jamming behavior.  


\subsection{Change of the jamming transition density under shear}
Next, we show that the jamming transition density shifts with the
application of shear. This is key to understanding the complicated
mechanical responses observed above. To obtain $\phi_{\rm J}(\gamma)$,
we firstly apply a shear strain $\gamma$ to the configuration at
$\phi=0.843$, trained at a particular depth $\phi_{\rm MAX}$. $\phi$ is
changed in $\Delta\phi=10^{-4}$ steps; when potential energy $U \sim
10^{-16}$, the corresponding $\phi$ is defined to be the jamming
transition density $\phi_J(\gamma)$ for a particular
$\gamma$. Figure~\ref{fig:fig5}~(a) shows the jamming transition density
$\phi_{\rm J}(\gamma)$ for different $\phi_{\rm MAX}$. In the small
$\gamma$ regime, $\phi_{\rm J}(\gamma)$ satisfies $\phi_{\rm
J}(\gamma)\sim \phi_{J0}$, whereas for larger shear strain
{\it i.e.}, $\gamma>0.01$, $\phi_{\rm J}(\gamma)$ deviates from $\phi_{J0}$ and
converges to another characteristic density $\sim 0.8426$, which we call
$\phi_{\rm JS}$,  
the jamming transition point for sheared configurations.
We find that $\phi_{\rm JS}$ is very close to values obtained using the
AQS shear reported in literatures~\cite{Heussinger2009Phys.Rev.Lett.,Vagberg2016Phys.Rev.E,Lerner2012Proc.Natl.Acad.Sci.USA}.  
The change in $\phi_{\rm J}(\gamma)$ is thus attributed to the loss of memory of the initial configuration due to shear. 
This small upward shift of $\phi_{\rm J0}$ under shear has been reported in
several studies~\cite{Heussinger2009Phys.Rev.Lett.,Vagberg2016Phys.Rev.E,Zheng2018Chin.Phys.B}. 
We address that 
this small shift in $\varphi_{\rm J}$ is responsible for both shear jamming and shear melting.
The same results are also obtained using constant pressure simulation by
applying steady shear (see Supplementary Information for simulation
details~\cite{zotero-12580}). The characteristic density obtained at a
constant low pressure, {\it i.e.}, $P \lesssim 10^{-5}$,  is equivalent to the jamming transition density. Figure~\ref{fig:fig5} (b) shows the density at constant pressure $P = 10^{-5}$ when a shear strain $\gamma$ is applied to configurations trained with various $\phi_{\rm MAX}$. The obtained densities $\phi(\gamma, P=10^{-5})$ show the same behavior as $\phi_{\rm J}(\gamma)$.
\subsection{Critical behavior of static properties}
Next, we discuss the jamming criticality of static mechanical properties at $\gamma = 0$ such as pressure $P$, shear modulus $G$, and potential energy $U$ for various $\phi_{\rm MAX}$. Figure~\ref{fig:fig6} (a) shows how pressure $P$ varies with $\phi-\phi_{\rm J0}$. We see that $P$ satisfies $P\sim \phi-\phi_{\rm J0}$ for all $\phi_{\rm MAX}$. Note that we present $\phi_{\rm J0}(\phi_{\rm MAX})$ just as $\phi_{\rm J0}$. Figure~\ref{fig:fig6} (b) shows how the shear modulus $G$ varies with $\phi-\phi_{\rm J0}$. We find that $G$ satisfies $G\sim (\phi-\phi_{\rm J0})^{0.5}$ for all $\phi_{\rm MAX}$. Finally, Figure~\ref{fig:fig6} (c) shows how the potential energy $U$ varies with $\phi-\phi_{\rm J0}$. We find that $U$ satisfies $U\sim \delta (\phi-\phi_{\rm J0})^{2}$ for all $\phi_{\rm MAX}$. In summary, critical scaling is successfully obtained for static properties regardless of the training history of the configuration ($\phi_{\rm MAX}$) as long as we set the jamming transition density to be $\phi_{\rm J0}$.

\begin{figure}
\includegraphics[width=\linewidth]{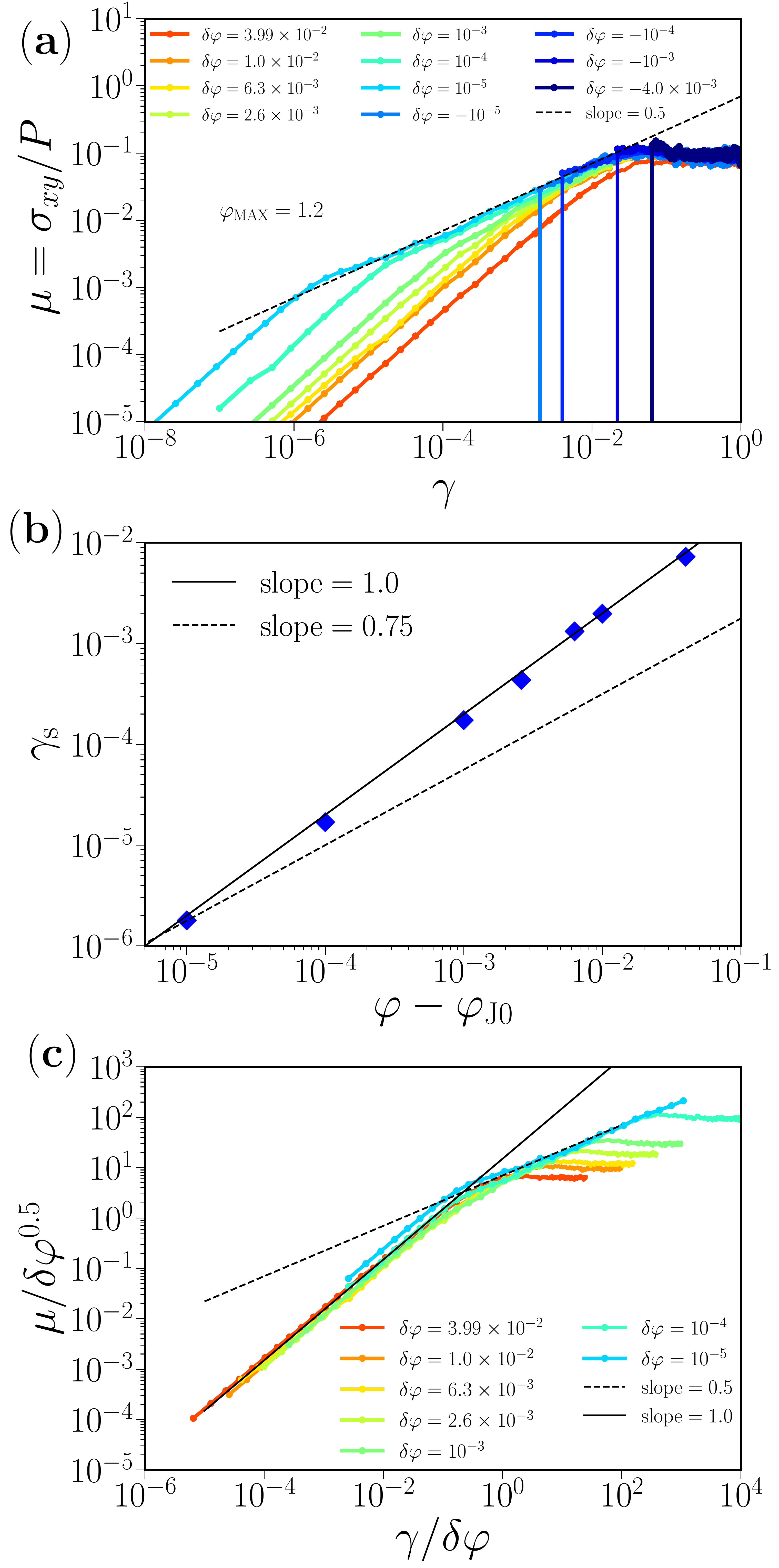}
\caption{Nonlinear rheology on softening of initial configurations trained at $\phi_{\rm MAX}=1.2$. (a) Friction coefficient $\mu=\sigma_{xy}/P$ vs $\gamma$ for various $\delta \phi= \phi-\phi_{\rm J0}$. There are three characteristic regimes: (i) $\gamma>0.01$, a yielded regime where $\mu$ is constant; (ii) intermediate $\gamma$, a softening regime where the friction coefficient obeys $\mu\propto \gamma^{0.5}$; (iii) small $\gamma$, an elastic regime where the friction coefficient obeys $\mu \propto \gamma$. (b) The crossover shear strain $\gamma_{\rm s}$ between the elastic and softening regimes as a function of proximity to the jamming transition density $\phi_{\rm J0}$. This characteristic shear strain is found to follow $\gamma_{\rm s}\sim (\phi-\phi_{\rm J0})$. (c) $\mu/\delta \phi^{0.5}$ vs $\delta \phi$ for the data shown in (a), where $\delta \phi=\phi-\phi_{\rm J0}$. Note the collapse of the data.
\label{fig:fig9}}
\end{figure}

\subsection{Critical behavior of the yield stress}
We go on to consider the critical behavior of the yield stress $\sigma^{\rm Y}$. In the present study, $\sigma^{\rm Y}$ is defined to be the average shear stress for large amplitudes of shear strain, $\gamma>0.2$, where the shear stress becomes nearly constant. In Fig.~\ref{fig:fig7}~(a), we show $\sigma^{\rm Y}$ as a function of $\phi-\phi_{\rm J0}$ for different $\phi_{\rm MAX}$. Note that $\phi_{\rm J0}$ depends on $\phi_{\rm MAX}$: critical scaling does not describe the relationship between $\sigma^{\rm Y}$ and $\phi-\phi_{\rm J0}$. To recover criticality for configurations trained with different $\phi_{\rm MAX}$, we adopt the jamming transition density $\phi_{\rm JS}$ instead of $\phi_{\rm J0}$; we immediately see a scaling relation $\sigma^{\rm Y}=A(\phi-\phi_{\rm JS})^\alpha$ with exponent $\alpha=1.01$ as shown in Fig.~\ref{fig:fig7}~(b). To understand this, we should note that the yield stress can only be obtained at large $\gamma$, where the memory of the initial configurations is lost. Hence, the corresponding jamming transition density should be $\phi_{\rm JS}$, the large $\gamma$ limit of $\phi_J(\gamma)$ shown in Fig.~\ref{fig:fig5}. 

A number of previous studies on the critical scaling of yield stress
have been carried out using finite shear rate
simulations~\cite{Tighe2010Phys.Rev.Lett.,Olsson2011Phys.Rev.E,Hatano2011J.Phys.:Conf.Ser.}. Here,
we discuss how our AQS simulations compare with finite shear rate
simulations. Figure~\ref{fig:fig7}~(c) shows data from
Ref.~\cite{Vagberg2016Phys.Rev.E}, the steady-state shear stress
obtained with finite shear rate simulation for various shear rates
$\dot{\gamma}$ and $\phi$. We compare these with the scaling of
$\sigma^{\rm Y}$ with $\phi-\phi_{\rm JS}$ obtained from our AQS
simulation. The asymptotic envelope of the finite shear rate simulation
data agrees with the AQS simulation data. This indicates that we have
successfully obtained $\sigma^{Y}$ in the AQS limit, {\it i.e.}, $\dot{\gamma}\to 0$. We reiterate that the critical exponent of yield stress is a topic of controversy, as described above. We are able resolve this; we find an exponent that is close to 1.0 using AQS simulations with an appropriate jamming transition density, $\phi_{\rm JS}$.

\subsection{Critical behavior of mechanical properties for a wide range of shear strain}
Until now, we have discussed two extreme cases for the strain, zero and
large $\gamma$. To bridge the two regimes, we investigate the critical behavior of the pressure $P$ and the shear stress $\sigma_{xy}$ for various $\gamma$ using initial configurations prepared with a training depth $\phi_{\rm MAX}=1.2$. In Fig.~\ref{fig:fig8}~(a), we plot $P(\gamma)$ as a function of $\phi-\phi_{\rm J}(\gamma)$. $\phi_{\rm J}(\gamma)$ is the jamming transition density for a particular $\gamma$ as shown in Fig.~\ref{fig:fig5}. We find that over a wide range of $\gamma$, the data collapses to $P(\gamma) \sim P_0(\phi-\phi_{\rm J}(\gamma))$. Hence, we conclude that $P(\gamma)$ is only governed by proximity to the jamming transition density $\phi_{\rm J}(\gamma)$ at a particular $\gamma$.

Moving on to shear stress when a strain $\gamma$ is applied,
Fig.~\ref{fig:fig6} shows that the yield stress, {\it i.e.}, 
the shear stress when large $\gamma$ is applied, exhibits critical behavior, $\sigma_{xy} \sim \phi-\phi_{\rm JS}$. When $\gamma$ is infinitesimally small, this relationship should obviously fail; one gets another relationship e.g. $\sigma_{xy}(\gamma) =  G/\gamma \sim (\phi-\phi_{\rm J}(\gamma))^{0.5}$. Therefore, we examine the range of $\gamma$ over which the relationship $\sigma_{xy} \sim \phi-\phi_{\rm J}(\gamma)$ is observed. In Fig.~\ref{fig:fig8}~(b), we plot the relationship between the shear stress $\sigma_{xy}(\gamma)$ and $\phi-\phi_{\rm J}(\gamma)$ for different $\gamma$ using configurations trained with $\phi_{\rm MAX}$=1.2. We confirm that for $\gamma \gtrsim 0.01$, all of the data collapses onto $\sigma_{xy}(\gamma) \sim \sigma_0(\phi-\phi_{\rm J}(\gamma))$. In summary, this analysis reveals that for large $\gamma>0.01$, $P$ and $\sigma$ obey the same critical scaling. This suggests that the ratio between $P$ and $\sigma_{xy}$, the friction coefficient $\mu=\sigma_{xy}/P$, is constant for different $\gamma$ and $\phi$. 

In Fig.~\ref{fig:fig9}~(a), we show the friction coefficient $\mu$ as a
 function of $\gamma$ at $\phi_{\rm MAX}=1.2$ for different $\delta
 \phi= \phi-\phi_{\rm J0}$. For $\delta \phi>0$, we find three
 characteristic regimes in $\mu$ as a function of $\gamma$: (i)
 $\gamma>0.01$, a yielded regime at large shear strain, where $\mu$ is
 constant, (ii) a softening regime at intermediate $\gamma$ where
 $\mu\propto \gamma^{0.5}$ and (iii) an elastic regime at small $\gamma$
 where $\mu \propto \gamma$. In (i), the shear stress obeys a critical
 scaling equivalent to the pressure,
{\it i.e.},  $\sigma_{xy} \sim \phi-\phi(\gamma)$. This is consistent
 with what we found for the critical scaling of the yield stress in
 Fig.~\ref{fig:fig6}~(a). In regime (ii), close to the jamming
 transition density, the softening behavior is now much clearer compared
 to the stress-strain curves, where it is partially hidden in the shear
 hardening region (see Fig~\ref{fig:fig2}). Moreover, we confirm that
 softening occurs even when $\delta \phi \lesssim 0$ {\it i.e.}, 
$\mu$ exhibits $\mu\propto \gamma^{0.5}$ for some $\gamma$ region at
 $\delta \phi=-10^{-4}$. This behavior is surprising and cannot be seen
 from the stress-strain curves because it is again masked by shear
 hardening. Thanks to the clear demarcation of a softening region in
 Figs~\ref {fig:fig9}~(a), we can estimate the crossover shear strain
 value $\gamma_{\rm s}$ between the elastic and softened regions. The
 characteristic shear strain is found to follow $\gamma_{\rm s}\sim
 (\phi-\phi_{\rm J0})^{1.0}$ as shown in Fig.~\ref{fig:fig9}~(b). 
The value of this critical exponent has been controversial. Two 
values of 0.75~\cite{Nakayama2016J.Stat.Mech.,Goodrich2016Proc.Natl.Acad.Sci.USA}
 and 
 1.0~\cite{Otsuki2014Phys.Rev.E,Boschan2016SoftMatter,Dagois-Bohy2017SoftMatter}
 have been reported so far.  
Obviously the exponent of 0.75 does not explain our data, as shown in Fig.~\ref{fig:fig9}~(b). 
The exponent obtained in the present study is clearly consistent with a value of 1.0.

We seek a scaling ansatz for $\mu$ vs $\gamma$ for the elastic and the softening regimes. In this approach, we assume that $\mu$ varies as 
\be
\mu(\delta \phi, \gamma)= \delta\phi^A {\cal F}(\gamma/\delta \phi^B), 
\ee 
where $\delta \phi=\phi-\phi_{\rm J0}$. The scaling function ${\cal F}(x)$ is proportional to $x$ when $x\ll1$ (the elastic regime), and otherwise 1/2 (softening). In the elastic regime, we obtain $\mu \propto \delta\phi^{A-B}\gamma$. Since $G\propto \delta \phi^{0.5}$ and  $P\propto \delta\phi$, we find $\mu \propto (G/P)\gamma\propto \delta \phi^{-0.5}\gamma$. Thus, in this regime, $A-B=-0.5$. In the softening regime, the scaling relation is $\mu \propto \delta\phi^{A-0.5B}\gamma^{0.5}$. Since $\mu(\delta \phi, \gamma)$ does not depend on $\delta \phi$, as shown in Fig.~\ref{fig:fig9}~(a), we get $A-0.5B=0$. Accordingly, we find $A=0.5$ and $B=1.0$. The exponent $B$ matches that of the elastic to softening crossover strain $\gamma_s$. We confirm the validity of this scaling ansatz by plotting $\mu / \delta \phi^{0.5}$ against $\gamma/\delta \phi$, as shown in Fig.~\ref{fig:fig9}~(c); both regimes follow the expected scaling. We note that there is a deviation in the large $\gamma$ region corresponding to the softening to yielding crossover.

\section{Summary}
We numerically simulate athermal particles under a quasi-static
shear. By employing the FIRE algorithm for energy minimization, we
create initial configurations with different depths of mechanical
training using a quasi-static cyclic volume deformations. We confirm
that $\varphi_{\rm J}$ varies with depth of mechanical training as
described in Fig~\ref{fig:fig1}. We then go on to change the density of
each jammed configuration, apply a uniform shear and consider the
mechanical response, as shown in Figs.~\ref{fig:fig2}--\ref{fig:fig4}. 
We observe either shear jamming or shear
melting. Notably, we find that the degree of mechanical training and
proximity to the jamming transition density strongly affect nonlinear
rheological response. We attribute this to a shift in the jamming
transition density under shear, as shown in Fig.~\ref{fig:fig4}, arising
from a loss of memory of the initial configuration induced by the shear,
with transition densities converging to a distinct jamming transition
density under shear, $\phi_{\rm JS}$. For a less annealed system, when
the packing fraction of the system $\phi$ satisfies $\phi_{\rm
J0}<\phi<\phi_{\rm JS}$, the system is initially jammed in response to a
small $\gamma$; when $\gamma$ is increased, it melts because the jamming
transition density also increases. This is the mechanism of shear
melting. On the other hand, with an intensively annealed system where
$\phi_{\rm JS}<\phi<\phi_{\rm J0}$, the system melts in response to a
small $\gamma$ applied to the system; when $\gamma$ increases, it jams
since the jamming transition density decreases. Thus, we reveal that a
shifted $\varphi_{\rm J}$ causes both shear jamming and shear melting.  

We also investigate the jamming criticality of both static
(Fig.~\ref{fig:fig6}) and dynamic quantities under shear
(Figs.~\ref{fig:fig7} and \ref{fig:fig8}). We show that the appropriate
critical density is equivalent to the jamming transition density at each
corresponding shear strain $\phi_{\rm J}(\gamma)$ as presented in
Fig.~\ref{fig:fig5}. Adopting this jamming transition density resolves
the controversy surrounding the critical scaling of the yield stress for
large shear
strains~\cite{Otsuki2009Phys.Rev.E,Tighe2010Phys.Rev.Lett.,Olsson2011Phys.Rev.E,Hatano2011J.Phys.:Conf.Ser.,Vagberg2016Phys.Rev.E,Heussinger2009Phys.Rev.Lett.,Heussinger2010SoftMatter}. 

We have also found that the crossover shear strain for elastic and softening regime is found to follow $\gamma_{\rm s}\sim (\phi-\phi_{\rm J0})^{1.0}$ as presented in Fig.~\ref{fig:fig9}~(b). Previous reports have shown that the critical exponent is controversial, varying from 0.75  to 1.0. Using the friction coefficient to disentangle softening and hardening, we obtain clearly separated softening behavior for a wide range of shear strain, revealing that the exponent is close to 1.0. We investigate nonlinear rheology near the jamming transition using the above friction coefficient, and find that softening occurs even below the jamming transition density, contrary to previous reports.

Finally, we remark that recent work~\cite{Heussinger2009Phys.Rev.Lett.,Vagberg2016Phys.Rev.E,Zheng2018Chin.Phys.B} has
shown that jamming configurations with a large applied shear exhibit a
slightly higher jamming transition density than the so-called isotropic
jamming transition density obtained from configurations without any
mechanical training. This is in agreement with our findings here, where
the jamming transition density as a function of shear strain $\phi_{\rm
J}(\gamma)$ for weakly trained systems, $\phi_{\rm MAX} \lesssim 0.845$,
increases with increasing $\gamma$. 
However, the essential underlying physics behind the small difference between the isotropic and anisotropic jamming transition densities have been overlooked.
Our findings provide a clear answer to the question over the small discrepancies between
the jamming transition point under/without shear and unified pictures how
the rich nonlinear behaviors of both shear melting and shear jamming by
tuning the amplitude or ``depth'' of the mechanical training. 

We thank S. Sastry, M. Otsuki, H. Hayakawa, K. Saitoh, Y. Jin, H. Yoshino, M. K. Nandi, M. Imamura, and T. Kurahashi for useful discussions. This work was financially supported by KAKENHI Grants 15H06263, 16H04025, 16H06018, and 19K03767.

%

\end{document}